# Deposition of $Cu_2Ta_4O_{12}$ by pulsed laser deposition

Andreas Heinrich[1], Bernd Renner[2], Robert Lux[1], Stefan G. Ebbinghaus[2], Armin Reller[2] and Bernd Stritzker[1]

[1] EPIV, University Augsburg, 86135 Augsburg, Germany
[2] FC I, University Augsburg, 86135 Augsburg, Germany

**Abstract**

$Cu_2Ta_4O_{12}$ (CTaO) thin films were successfully deposited on Si(100) substrates by pulsed-laser deposition technique. The crystalline structure and the surface morphology of the CTaO thin films were strongly affected by substrate temperature, oxygen pressure and target – substrate distance. In general during deposition of CTaO the formation of a $Ta_2O_5$ phase appeared, on which CTaO grew with different orientations.

We report on the experimental set-up, details for film deposition and the film properties determined by SEM, EDX and XRD.

## 1. Introduction

High dielectric constant materials allowing small capacitive elements are favourable for microelectronic devices like wireless communication systems such as cellular phones and global positioning systems. Recently, much attention has been paid to the perovskite related $CaCu_3Ti_4O_{12}$ with an apparent high dielectric constant in the order of $10^4$ for single crystals at room temperature for low frequencies. [1,2 and references therein]. This material has successfully been deposited on Si substrates [3,4]. The $AA'_3M_4O_{12}$-type of structure is characterized by a collective tilting of the oxygen octahedra along <111> direction, causing the A'-sites to have a rectangular four-fold co-ordination.

Another promising material might be $Cu_2Ta_4O_{12}$ (CTaO). Its crystalline structure can be derived from $CaCu_3Ti_4O_{12}$ by leaving the Ca postions and 1/3 of A' sites unoccupied. The structure is schematically shown in Fig. 1. Oxygen, copper atoms and $TaO_6$-octahedra are indicated. Single crystalline $Cu_2Ta_4O_{12}$ crystallizes in a slightly distorted pseudo-cubic structure with lattice constants of a=b=7.51 Å and c=7.52 Å [5].

The preparation of CTaO thin films on silicon subtrates by pulsed laser deposition is – to our knowledge – reported for the first time.

## 2. Experimental details

The experimental set-up is shown in Fig. 2. A LPX 300 (Lambda Physics) excimer laser (KrF) was used to generate 30ns UV-pulses (248nm). To compensate the divergence of the laser beam the light passes a lens with f=10m. A second lens in front of the chamber images the aperture on a rotating laser target. Thereby the cross section of the beam is reduced about 8 times and hence the energy density on the target is 5-6 J/cm$^2$. This results in an evaporation of the target material and a plasma plume perpendicular to the target surface. The shape and spread of the plasma plume depends on the target material, the laser energy and on the gas pressure. During the ablation proscess oxygen pressure and laser energy were adjusted to let the plasma plume just reached the sample.

For target preparation CuO and $Ta_2O_5$ was mixed and heated up to 1050°C for one day. This results in polycrystalline $Cu_2Ta_4O_{12}$ powder, which was pressed into a cylindrical pellet and sintered at 900°C for 6 hours. The target rod was fixed on a revolvable axis to allow an evenly degradation during the ablation process. In general (100) Si substrates were used, having a very small lattice mismatch of 2.4%. To remove the natural $SiO_2$ layer from the surface all substrates were etched in diluted hydrofluoric acid for 30 seconds.

## 3. Results and discussion

After film deposition the film thickness was first measured in dependence of laser energy. In a second step the influence of the oxygen pressure on the formation of CTaO was analyzed. Figure 3 shows the obtained film thickness as a function of laser energy. These experiments were performed at an oxygen pressure of $2*10^{-2}$ mbar and 800°C substrate temperature. The repetition rate of the laser beam was 8 Hz and 10.000 pulses were applied for each ablation.

The laser target was polished, then four ablations with different energy (400mJ, 600mJ, 800mJ and 965mJ) were carried out. The laser beam hit the same trace of the rotating target all the time. Afterwards the target was moved perpendicular to the beam, thus an unaffected area could be used for another four ablations with decreasing energy (1000mJ, 800mJ, 600mJ and 400mJ).

As one expects, film thickness increases with increasing energy. Another important result is the strong dependence of the growth rate on the history of the target. This can be seen very easily by comparing sample four and five. In both cases about 1000mJ/pulse were used. Sample five was achieved with a laser beam on a new trace of the target and sample four on a trace already used for 3 ablations. On sample five a higher film thickness (160 nm in stead of 100 nm) and thus a higher growth rate was achieved. This effect worsens as one applies lower laser energies (sample one - new trace and sample eight - trace already used for sample five, six and seven). No appreciable film thickness for sample eight was found.

A change also can be noticed optically: After a few thousand pulses the green colored target changed its color to brown within the area the laser beam had hit the rod. To ensure equal starting conditions for deposition the target was polished in advance of each ablation process.

To study the influence of oxygen pressure on the formation of $Cu_2Ta_4O_{12}$ four different oxygen pressures were used ($2*10^{-2}$ mbar, 0.5 mbar, 1 mbar and 1.6 mbar). In general 30.000 pulses were applied. All samples were investigated by XRD, SEM and EDX.

X-Ray diffraction patterns of films grown at $2*10^{-2}$ mbar showed no hint of a CTaO phase. All determined peaks belonged to $Ta_2O_5$. This result is supported by EDX measurements. Besides a Si peak of the substrate only Ta and O could be found. There were no traces of Cu found on the sample. Thus we increased the oxygen pressure to 0.5 mbar and reduced the distance target – substrate from 4.5 cm to 2.5 cm.

In this case the XRD-pattern showed a clear $Cu_2Ta_4O_{12}$ phase. But the dominant intensities were still due to $Ta_2O_5$. SEM and EDX revealed a $Ta_2O_5$ layer with some small CTaO areas. An X-Ray diffraction spectra of a sample deposited at an oxygen pressure of 1.0 mbar is shown in Fig. 4. There are still some $Ta_2O_5$ phases, but clearly CTaO in different orientation as well. An SEM overview of the sample is shown in Fig. 5. The image was taken near the edge of the specimen. Towards the center (right side) a closed layer is observable. As one expects the film thickness decreases towards the edge of the specimen due to the dimensions of the plasma plume. This results in more and more disconnected islands. To analyze the difference between holes and islands/film two EDX spectra were taken, as indicated in Fig. 6 (no. 1: hole; no. 2: layer). In both cases Ta and O were found, but in contradiction to position 2, position 1 did not show any hint of Cu. Thus and with the XRD data we conclude that the crystals are CTaO and the bottom refers to $Ta_2O_5$.

A closer look on the sample is given in Fig. 7. It is shown that CTaO grows in different directions on the $Ta_2O_5$ – layer. The grown film consists of hundred nm long plate like crystals with a thickness of about 10nm. One can clearly observe areas with crystals growing parallel to each other. An intergrowth of two or more of these areas with different orientations can result in bulges as can be seen in Fig. 5.

To complete the study of the influence of oxygen pressure on film formation the pressure was increased to 1.6 mbar. The XRD-pattern (Fig. 8) clearly revealed CTaO with different orientations. In comparison to an oxygen pressure of 1.0mbar the CTaO is more dominant than $Ta_2O_5$. This assumption is confirmed by SEM images, taken from the edge of the specimen, shown in Fig. 9. Three regions are indicated, which were examined by EDX. In area 1 only Ta and O were found indicating that this is again the bottom $Ta_2O_5$-layer. Position 2 and 3 exhibited a strong copper content. These two areas refer to the CTaO-layer. Thereby the crystals in position 3 are much more dominant. Different film growth on the substrate might be due to different orientations of the forming $Ta_2O_5$-layer. A stronger CTaO growth can be expected on $Ta_2O_5$ crystals which are oriented in a way to provide a lower lattice misfit. These could be the areas with strong crystal growth as it can be observed in area 3.

In comparison to an oxygen pressure of 1.0 mbar we received at 1.6 mbar a CTaO growth even in the "holes". This may be due to the urge of $Cu^{2+}$-ions to be reduced at elevated temperatures even at ambient oxygen pressure. Resulting $Cu^{1+}$ or Cu metal is not able to participate in the formation of $Cu_2Ta_4O_{12}$.

## 4. Conclusion

We showed that it is possible to grow $Cu_2Ta_4O_{12}$ on Si substrates. Due to the laserimpact on the target, the target's structure changes drastically. Thus it is necessary to polish it in advance of each ablation.

The oxygen pressure should be at least 1.0 mbar to gain the $Cu_2Ta_4O_{12}$ phase. We could show, that a $Ta_2O_5$ layer between $Cu_2Ta_4O_{12}$ and Si is forming. The $Cu_2Ta_4O_{12}$ crystals grew in different orientations, preferably in certain areas on the $Ta_2O_5$ layer, which might have favorable crystal orientation.

The dielectric properties of films grown in this way is subject of actual studies. Additionally we have started to epitaxially grow $Cu_2Ta_4O_{12}$ thin films on different substrates. Results will be published in the near future.

**List of figures and tables**

**Figures**

Fig. 1 :

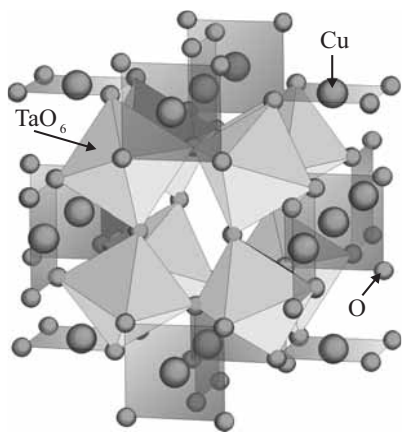

Fig. 2:

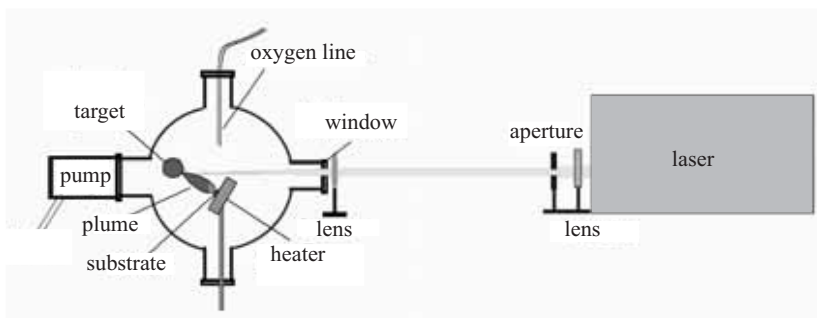

Fig. 3:

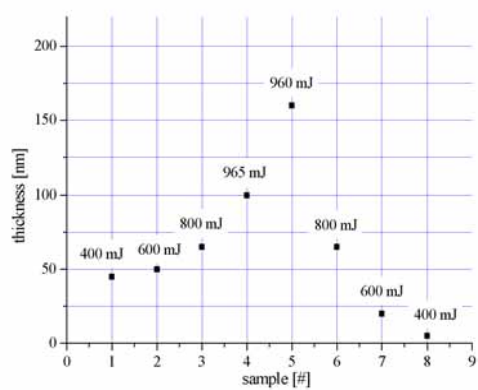

Fig. 4:

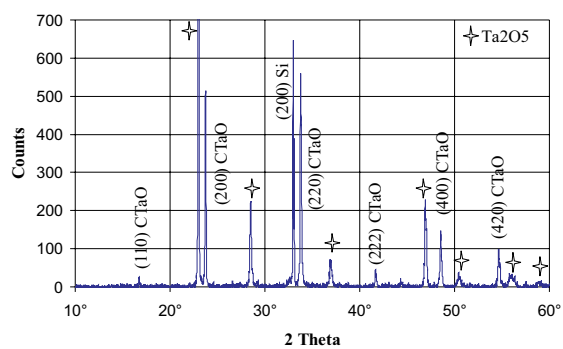

Fig. 5

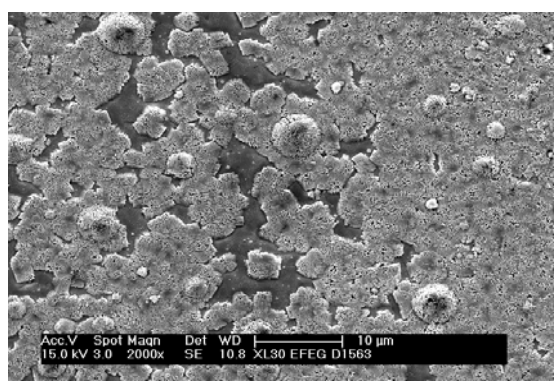

Fig. 6

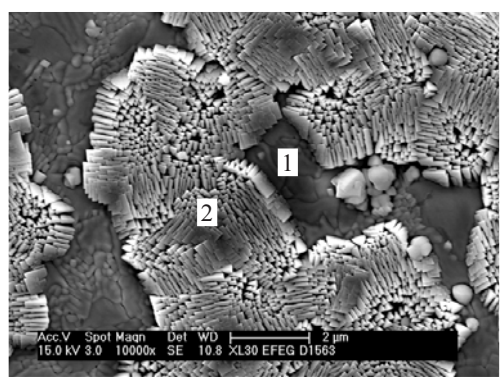

Fig. 7:

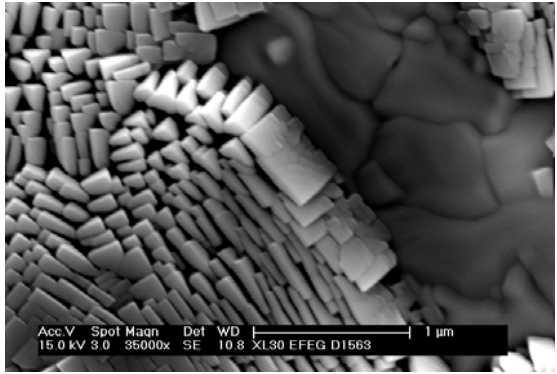

Fig. 8:

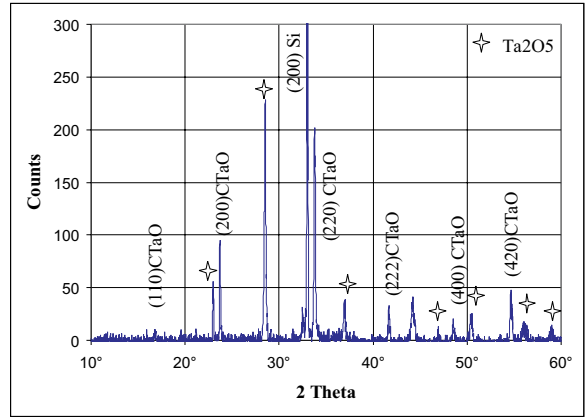

Fig. 9:

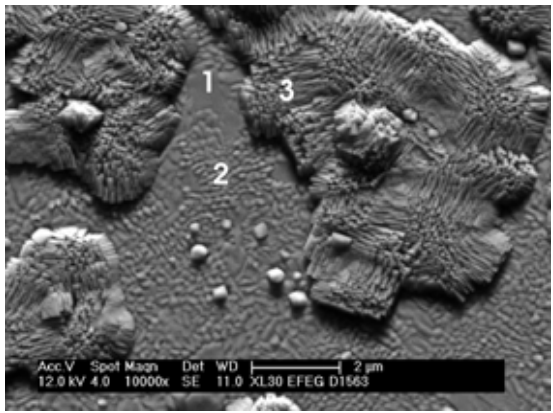